\documentclass[%
 reprint,
 amsmath,amssymb,
 aps,
 pra,
]{revtex4-1}

\usepackage{graphicx}
\usepackage{dcolumn}
\usepackage{bm}
\usepackage{natbib}
\usepackage{siunitx}
\usepackage{textcomp}
\usepackage{color}
\usepackage{tikz}
\usetikzlibrary{shapes}
\usepackage{hyperref}

\begin{document}

\preprint{APS/123-QED}

\title{Coupling two laser-cooled ions via a room-temperature conductor}

\author{Da An}
\author{Alberto M. Alonso}
\author{Clemens Matthiesen}
\author{Hartmut H\"{a}ffner}

\affiliation{
 Department of Physics, University of California, Berkeley, CA 94720
}%

\date{\today}

\begin{abstract}
We demonstrate coupling between the motions of two independently trapped ions with a separation distance of $620$~{\textmu}m. The ion-ion interaction is enhanced via a room-temperature electrically floating metallic wire which connects two surface traps. Tuning the motion of both ions into resonance, we show flow of energy with a coupling rate of $11$~Hz. Quantum-coherent coupling is hindered by strong surface electric-field noise in our device.
Our ion-wire-ion system demonstrates that room-temperature conductors can be used to mediate and tune interactions between independently trapped charges over distances beyond those achievable with free-space dipole-dipole coupling. This technology may be used to sympathetically cool or entangle remotely trapped charges and enable coupling between disparate physical systems.
\end{abstract}

\pacs{Valid PACS appear here}
\maketitle

Trapped-ion systems offer a unique platform for precision measurements of fundamental constants and searches for new physics~\cite{Devlin2021, Pruttivarasin2015a}. Trapped ions are also one of the leading qubit candidates for quantum information processing, with demonstrated high-fidelity gate operations~\cite{Gaebler2016, Ballance2016, Srinivas2021} and fully-controllable small quantum computers~\cite{Wright2019a, Pogorelov2021, Pino2021}. Large-scale designs may be achievable with a modular architecture consisting of several quantum registers linked together with quantum interconnects~\cite{Brown2016}. These interconnects may involve shuttling ion qubits between sites~\cite{Kielpinski2002CCD, Kaufmann2017-swapping, Pino2021} or establishing photonic links~\cite{Olmschenk2009-teleportation,Stephenson2020-entanglement}. 
Both approaches come with their own technical challenges, however. Shuttling, for instance, requires dynamic and precise control of the trapping potential, and complex trap layouts with many individually addressed electrodes, while coupling rates in photonic schemes are limited by photon collection and detection efficiencies as well as the probabilistic nature of entanglement generation.

\begin{figure}[t!]
    \centering
    \includegraphics[scale=1]{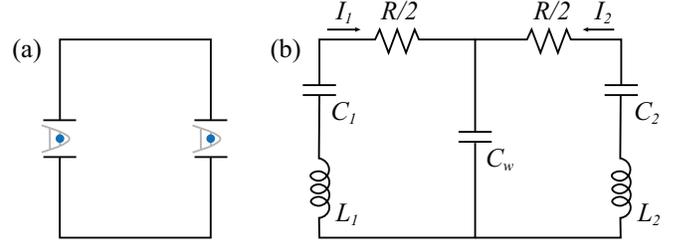}
    \caption{(a) Schematic of two harmonically bound ions connected with a wire. (b) Equivalent circuit model of two ions coupled with a wire. The ion motion is modeled as inductance $L_i$ and capacitance $C_i$ while the current $I_i$ corresponds to the ion velocity~\cite{Wineland1975}. The coupling wire has ohmic resistance $R$ and capacitance $C_w$ to the rest of the trap.}
    \label{fig:equiv_circuit}
\end{figure}

Here, we explore an alternative interconnect mechanism: coupling of two remote ions in separate traps mediated via an electrically floating metallic wire. First envisioned by Heizen and Wineland~\cite{Heinzen1990}, this scheme relies on coupling the motional dipole of each ion to a conductor wire located between them. As each trapped ion oscillates, its motion induces oscillating image charges in the shared wire~\cite{Wineland1975}, which acts as an intermediate bus to transfer information between the ions. A schematic of this system is illustrated in Fig.~\ref{fig:equiv_circuit}(a). The original idea of electrical coupling has since been demonstrated between resistively cooled ions in separate Penning traps~\cite{Tu2021Tank-CircuitCooling}, and conceptually adapted from Penning to rf Paul traps in the context of quantum information~\cite{Daniilidis2009}.

\begin{figure}[t!]
    \centering
    \includegraphics[scale=0.38]{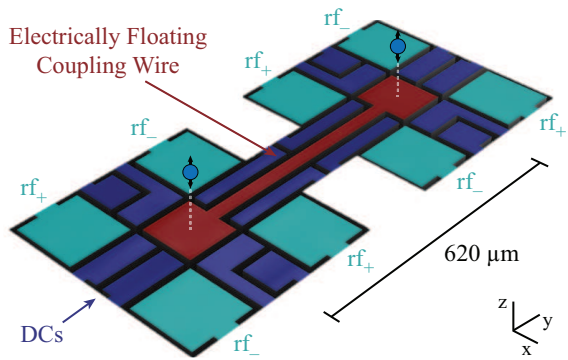}
    \caption{The dual surface-electrode trap consists of two trapping sites, each composed of 4 rf electrodes (cyan) driven out-of-phase and 8 independently controlled DC electrodes (blue). The two sites are linked with an electrically floating wire (red) which is engineered to maximize ion-wire coupling. Ions are separated by $620$~{\textmu}m and the vertical motion of each ion induces image currents in the wire and vice-versa.} 
    \label{fig:trap}
\end{figure}

In our experiments, the remote interaction between laser cooled ions is achieved in a novel surface ion trap geometry with an integrated coupling wire, see Fig.~\ref{fig:trap}. 
We show that two ions independently trapped at opposite ends of the wire interact when their motions are tuned into resonance.
The interaction signatures manifest as energy transfer between the ions.

Wire-mediated coupling may be compared to direct dipole-dipole coupling, which was achieved by placing ions in separate potential wells of the same surface trap at separation distances of approximately $50$~{\textmu}m~\cite{Harlander2011, Brown2011}.
The direct Coulomb interaction strength, $\Omega_\mathrm{ex}$, falls off as $1/r^3$, where $r$ is the ion-ion separation, limiting practical applications, whereas wire-mediated coupling scales as $1/r$~\cite{Daniilidis2009}. With our experimental parameters ($r=620$~{\textmu}m), we expect a 60-fold improvement in the coupling rate. 

Our demonstration of inter-trap coupling and energy exchange promises direct applications for sympathetic cooling and precision measurements of charged particles -- in particular for particles not amenable to laser cooling, such as protons~\cite{Bohman2018,Tu2021Tank-CircuitCooling}, electrons~\cite{Matthiesen2021-electron-trapping}, or their antiparticles.
This may also open up applications for trapped-ion/electron quantum computing~\cite{Daniilidis2009, Daniilidis2013electron} and hybrid quantum interconnects linking trapped ions to solid-state quantum systems~\cite{Daniilidis2013a,Kotler2017, Eltony2016}. 


The ion-wire-ion system can be described as a pair of coupled harmonic oscillators, with the approximate Hamiltonian
\begin{equation} \label{eq:classicalHO}
\hat{\mathcal{H}} = \sum_{i=1}^2 \frac{p_i^2}{2 m_i} + \sum_{i=1}^2 \frac{1}{2} m_i \omega_i^2 x_i^2 + 2 \kappa \prod_{i=1}^2 \sqrt{m_i \omega_i} x_i,
\end{equation}
where $m_i, \omega_i, p_i, x_i$ are the mass, frequency, momentum, and position of ion $i$, respectively, and $\kappa$ represents the coupling strength.

Using an electrically equivalent circuit of the interaction~\cite{Wineland1975, Heinzen1990, Daniilidis2009}, we provide an intuitive framework to understand the ion coupling in terms of the trap geometry. A diagram of the equivalent circuit is shown in Fig.~\ref{fig:equiv_circuit}(b).
Here, the harmonic motion of two ions coupled by a wire is represented by two series $RLC$ circuits with a shunt capacitor~$C_{w}$, such that
\begin{equation} \label{eq:eom_ion_circuit}
L_i = \frac{m D_\mathrm{eff}^2}{q^2} ~ , ~ C_i = \frac{1}{\omega_t^2 L_i}
\end{equation}
for ions $i = 1 ,2$ of charge $q$ and mass $m$, where $D_\mathrm{eff} \equiv U_w / E^w$ is the effective distance between the ion and the coupling wire and determined by the ratio of the wire voltage, $U_w$, to the generated electric-field at the trapping position, $E^w$~\cite{Wineland1975, Heinzen1990}. While $D_\mathrm{eff}$ is proportional to the physical ion-wire distance, its exact value depends on the geometry of the wire.
The current, $I_i$, in each circuit branch corresponds to the velocity of ion $i$.

Using this picture, the coupling strength, $\kappa$, under resonant energy exchange of the wire-mediated coupling may be expressed in terms of the circuit parameters as
\begin{equation} \label{eq:equiv_circuit_g}
\kappa = \frac{\pi}{2} \frac{1}{\sqrt{L_i C_w}} = \frac{\pi}{2} \frac{q^2}{m \omega} \frac{1}{C_w D^2_\mathrm{eff}}.
\end{equation}
In this form, we find that the coupling strength increases with the magnitude of the electric dipole moment of the motion of each ion, favoring particles with high charge to mass ratios (or multiple trapped charges) and low trap frequencies. In addition, the coupling strength benefits from wire designs that minimize $C_w$ and $D_\mathrm{eff}$, as the capacitance tends to short out the coupling signal and small ion-wire distances increase the induced image charge.

Positioning a single wire above a linear surface ion trap, as proposed in Ref.~\cite{Daniilidis2009}, minimizes the capacitance $C_w$ by virtue of being far away from other electrodes, but limits the effective ion-electrode distance and poses other practical challenges, such as precise integration and maintaining good optical access to the ion. We address these challenges with a unique dual ion trap design, see Fig.~\ref{fig:trap}, where the coupling wire connects two independent 4-rf-electrode traps~\cite{An2018}, and replaces the center electrode for both.

The surface-integrated wire features two square paddles connected by a narrow strip. The paddle centers, which coincide with the in-plane trap centers, are separated by $620$~{\textmu}m. The paddle geometry, with dimensions $120$~{\textmu}m $\times$ $120$~{\textmu}m, optimizes ion-wire coupling for the vertical mode at each trapping site for a trapping height of $50$~{\textmu}m. Boundary-element method calculations at this physical ion height reveal an effective ion-wire distance of $D_\mathrm{eff} = 130$~{\textmu}m for our wire geometry. 
Trap electrodes, defined by a $1$-{\textmu}m thick aluminum-copper film, are separated with $15$-{\textmu}m wide trenches. Finite-element analysis of the paddles and the connecting $30$-{\textmu}m wide strip yield a total capacitance of the wire, $C_w$, of $30$~fF. Details on the trap fabrication are described in Ref.~\cite{An2018}.

There are two key reasons for using the 4-rf-electrode geometry for this ion-ion coupling experiment. First, as the vertical ion mode couples more strongly to the wire than the transverse modes of motion, it should be used for ion-ion interactions. Since the interaction is of resonant nature, precise control and high stability of the vertical trapping potential are crucial. This is easier to achieve when the confinement can be controlled with DC instead of AC fields. The 4-rf-electrode trap, when driven out of phase, features an rf null normal to the paddle center~\cite{An2018} and the vertical potential is controlled purely with DC fields as desired.
Second, maintaining well-defined harmonic trapping potentials at the two trapping sites requires low cross-talk between trapping fields. In particular, pick-up of the rf drive on the floating wire could disturb the potentials significantly. With the 4-rf-electrode trap driven in the out-of-phase configuration, the rf fields average out along the electrically floating wire, minimizing potential rf pickup.

In the experiments discussed here, a single $^{40}$Ca$^+$ ion is held in each trap, with planar trap frequencies $\omega_{x,y} \approx 2 \pi \times3$~MHz and typical vertical trap frequencies of $\omega_z = 2 \pi \times 2$~MHz. Confinement in the planar directions is generated with an out-of-phase rf drive at $36$~MHz and approximately $100$~$\mathrm{V}_\mathrm{pp}$. The two traps are driven with the same rf source. 
For an ion height of 60~{\textmu}m and a resonant trap frequency $\omega_z = 2 \pi \times 2$~MHz, we expect a coupling rate of $\kappa \approx 2 \pi \times 10$~Hz.

We can measure this energy exchange by engineering a temperature difference, $\Delta T = T_2 - T_1$, between the vertical modes of \textsc{Ion 1} and \textsc{Ion 2}, and tuning their frequencies into resonance. For an isolated ion-wire-ion system, we expect $T_1$ and $T_2$ to perfectly swap as a function of time. In practice, we find this picture is strongly modified by electric field noise generated by the trap electrode surfaces, likely acerbated by the coupling wire: the electrically floating nature of the wire makes it susceptible to charging and we observe manifestations of dynamic charging at both slow and fast timescales. Motional heating rates probe electric field noise at the trap frequency (fast noise) and we measure heating rates on the order of $100$~quanta/ms for the vertical mode. This is two orders of magnitude higher than expected based on measurements of a different trap on the same chip~\cite{An2019-distance-scaling}.
While the exact noise mechanism is unclear, we suspect noise caused by imperfect shielding of the dielectric layer or fluctuating excess surface charge carriers~\cite{Teller2021}.

In addition to the fast charge fluctuations, the total number of charges determines the potential of the coupling wire, and a voltage offset from ground perturbs the trapping potential. Large unknown voltage offsets in excess of 1\,V prevent us from trapping ions. However, we are able to steer the wire charge towards known values using the photoelectric effect~\cite{Wang2011Laser-inducedTraps, Harlander2010}. Specifically, exposure of $375$~nm light directly onto the wire excites the transfer of electrons from the wire into the trap ground, thus increasing the net charge on the wire. The opposite effect may be achieved by shining the same light onto electrodes that neighbor the floating wire. In this way, we may regulate the wire charge towards neutral, minimizing its effect on the vertical trapping potential.

The strong electric-field noise in our system limits motional coherence, washing out the signatures of coherent energy exchange and preventing cooling of the ions into their motional ground states. As such we cannot measure wire-mediated energy exchange at the level of single quanta. In light of these limitations, we instead demonstrate the wire-mediated coupling through measurements of sympathetic heating and sympathetic reduction of ion temperature.

The experiments are performed with a single $^{40}$Ca$^+$ ion held in each zone of the dual trap. Each ion is addressed with an independent set of lasers required for cooling and state manipulation.
Excessive photoelectric charging of the trap surface prevents cooling of the vertical ion motion with a conventional $397$~nm beam oriented along the vertical axis. Instead, Doppler cooling is achieved on the $P_{1/2} \leftrightarrow D_{3/2}$ transition using a red-detuned vertical infrared $866$~nm beam, which does not generate any measurable photoelectric effect.
A separate $397$~nm beam parallel to the trap surface maintains Doppler cooling of the planar modes and allows for fluorescent detection of each ion onto separate photomultiplier tubes.
State manipulation is performed by two individually focused $729$~nm beams which address the $S_{1/2} \leftrightarrow D_{5/2}$ qubit transitions and point along the vertical modes of each ion. We perform temperature measurements by extracting the mean motional occupation number, $\Bar{n}$, from fits of Rabi oscillations of the qubit transition~\cite{Wineland1998-experimental-issues}. 

\begin{figure}[t!]
    \centering
    \includegraphics[scale=1]{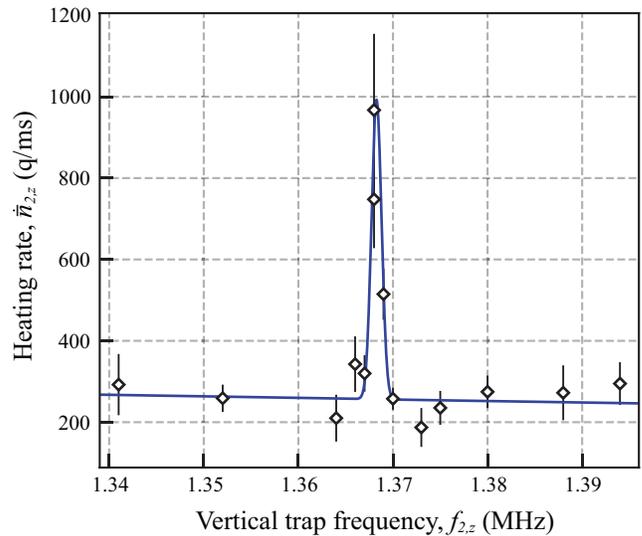}
    \caption{Heating rate ($\Dot{\Bar{n}}_{2,z}$) spectroscopy for the vertical mode of \textsc{Ion 2} in the presence of \textsc{Ion 1}, which is held at $\Bar{n}_{1,z} \gtrsim 10,000$ quanta and $\omega_{1,z} = 2 \pi \times 1.3680(6)$~MHz. Error bars represent one standard deviation. Near the resonance condition, \textsc{Ion 2} is sympathetically heated by \textsc{Ion 1}. The blue curve represents a Gaussian fit ($527 \pm 160$~Hz width) taking into account also the observed $1/f$ frequency scaling of background electric-field noise.}
    \label{fig:coupling_resonance}
\end{figure}


In our experiment, \textsc{Ion 1} is trapped $60$~{\textmu}m above the surface and initialised to a vertical mode temperature of $\Bar{n}_{1,z} \gtrsim 10,000$ quanta for frequency $\omega_{1,z} = 2 \pi \times 1.3680(6)$~MHz.
\textsc{Ion 2} is located $80$~{\textmu}m above the opposite end of the coupling wire and initialized to $\Bar{n}_{2,z} = 200$ quanta.
Figure~\ref{fig:coupling_resonance} shows the heating rate of \textsc{Ion 2} as a function of its motional frequency.
When the resonance condition of $\omega_{2,z} = \omega_{1,z}$ is established, we find the heating rate spikes from a baseline of $250$~quanta/ms to approximately 1000~quanta/ms. The slight decrease of the baseline heating rate over the measured frequency range is consistent with the observed $1/f$ surface electric-field noise~\cite{Brownnutt2015}.

The width of the sympathetic resonance is related to the mutual stability of the vertical motion in the two traps. Our data in Fig.~\ref{fig:coupling_resonance} suggests relative trap frequency fluctuations of the vertical mode of less than $530 \pm 160$~Hz during the few-second-long measurement. Independent measurements indeed verify that the stability of the vertical frequency is better than 1\,kHz. We expect the motional linewidth to be broadened by anharmonicities of the trap potential in combination with the large ion temperatures.


\begin{figure}[t!]
    \centering
    \includegraphics[scale=0.95]{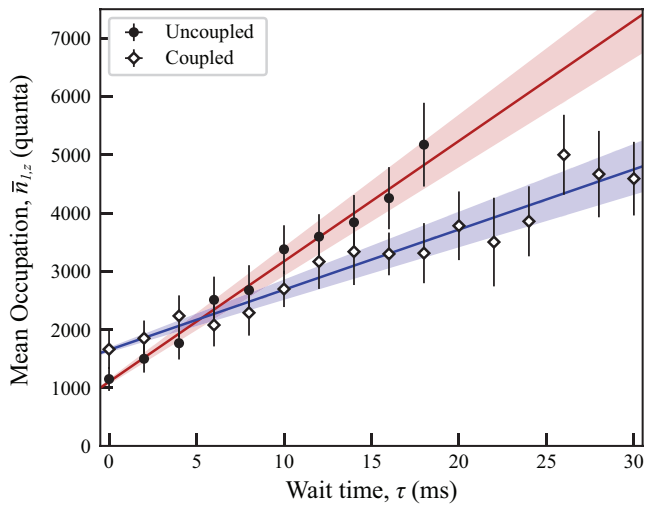}
    \caption{Mean motional occupation ($\Bar{n}_{1,z}$) of \textsc{Ion 1} is plotted with error bars representing one standard deviation. The baseline $\Bar{n}_{1,z}$ heating of an uncoupled \textsc{Ion 1} is plotted as filled circles, along with a heating rate fit of $206(20)$~quanta/ms (red). In this case, the initial $\Bar{n}_{1,z}$ is arbitrary and does not affect the heating rate. A second experiment, shown with data plotted as unfilled diamonds, demonstrates sympathetic reduction of \textsc{Ion 1}'s temperature induced by resonant wire-mediated coupling with \textsc{Ion 2} held at $182(15)$~quanta. The resulting reduced heating rate is calculated to be $102(12)$~quanta/ms (blue).}
    \label{fig:coupling_cooling}
\end{figure}

Next, we aim to quantify the coupling rate between the ions. For this measurement we shuttle the ions to lower trapping heights which increases the coupling signal: \textsc{Ion~1} (the measurement ion) is trapped at $50$~{\textmu}m height and \textsc{Ion~2} (the auxiliary ion) at $70$~{\textmu}m.
Bringing the ions closer to the surface increases their exposure to surface electric-field noise, so at the same time we increase the resonance frequency $\omega_z$ from $2 \pi \times 1.368$~MHz to $2 \pi \times 1.990$~MHz. This lowers noise by roughly a factor of 2 while reducing the coupling rate by a factor of $\sqrt{2}$. The magnitude of electric-field noise precludes any observations of temperature oscillations due to the wire-mediated energy exchange between the remote ions. However, we can still monitor the temperature evolution during resonant coupling under these conditions and extract a coupling rate.

The experiment consists of two parts. First, we extract a baseline heating rate for \textsc{Ion~1} in the absence of coupling to a second ion. The data in Fig.~\ref{fig:coupling_cooling} (filled circles) give an uncoupled heating rate of $\Dot{\Bar{n}}_{u} = 210(20)$~quanta/ms at $\omega_{1,z} = 2 \pi \times 1.990$~MHz when \textsc{Ion~1} is trapped at $50$~{\textmu}m height and the second trap is empty.
In the second part, we work with one ion in each trap. \textsc{Ion~1} is initialized to a high temperature of $\Bar{n}_{1,z} \approx 1000$~quanta, and then left sans laser light until the measurement. \textsc{Ion~2} is continuously Doppler cooled to a temperature $\Bar{n}_{2,z} = 182(15)$~quanta. Its trap frequency is tuned to be resonant with \textsc{Ion~1} for the entire duration of the experiment. On resonance, the cooled \textsc{Ion~2} is expected to sympathetically cool the hot \textsc{Ion~1}. Under these conditions, we record the time evolution of $\Bar{n}_{1,z}$, shown in Fig.~\ref{fig:coupling_cooling} as unfilled diamonds.
While we do not strictly observe sympathetic cooling, that is, the temperature of \textsc{Ion~1} is not decreasing in time, we find that the coupling sympathetically reduces the heating rate of \textsc{Ion~1} to $\Dot{\Bar{n}}_{c} = 102(12)$~quanta/ms, see blue fit in Fig.~\ref{fig:coupling_cooling}. For long wait times, this coupling effect leads to lower \textsc{Ion~1} temperatures compared to the uncoupled case.

Using a classical model of the ion motions, we extract an effective coupling rate of $\kappa = (\Dot{\Bar{n}}_{u} - \Dot{\Bar{n}}_{c}) / (\Bar{n}_{1,z} - \Bar{n}_{2,z}) = 2 \pi \times 11.1$~Hz from the data in Fig.~\ref{fig:coupling_cooling}, where $\Bar{n}_{1,z}$ and $\Bar{n}_{2,z}$ are the mean phonon numbers for \textsc{Ion~1} and \textsc{Ion~2}.
The difference between our measured coupling rate and the predicted coupling rate of $\kappa = 2 \pi \times 10.2$~Hz
may be due to uncertainties in the simulated wire capacitance, $C_w$, or uncertainties in the ion heights. As a comparison, this wire-mediated coupling strength is about 60 times faster than the expected bare Coulomb coupling strength
at a $2 \pi \times 1.99$~MHz resonant interaction frequency for the same ion-ion separation. 

In conclusion, we have established coupling between two remotely trapped ions enhanced via a solid-state wire at room temperature.
In principle, the wire-mediated coupling design opens new avenues to cool, detect, or otherwise couple remotely trapped charged particles such as antiprotons~\cite{Niemann2019,Tu2021Tank-CircuitCooling} or electrons~\cite{Zurita-Sanchez2008, Matthiesen2021-electron-trapping}. In fact, this scheme is particularly attractive for electrons, as the coupling strength increases by two orders of magnitude when replacing $^{40}$Ca$^+$ ions with trapped electrons at $100$~MHz trap frequency.
Applying our proof-of-concept demonstration as a tool for sympathetic cooling in spectroscopy or quantum science schemes will require improvements in the experimental setup, however. Specifically, the electric-field noise in our device is currently too large for most practical applications.
Some scenarios only require a moderate improvement to compete with existing approaches, though. For instance, cooling a remotely trapped ion to approximately $100$~mK in a device identical to ours -- which would otherwise require a dilution refrigerator -- should be possible with an order of magnitude reduction in heating rates.

In the context of using ion-wire-ion coupling for quantum coherent energy exchange of $^{40}$Ca$^+$ ions, we must reduce the system noise by at least three orders of magnitude. Such a task is onerous even with the demonstrated toolkit of techniques for surface ion-trap noise reduction, such as Ar$^+$ bombardment~\cite{Hite2012, Daniilidis2014-ion-milling} and operation at cryogenic temperatures~\cite{Brandl2016-cryogenic-trapping}. 

To be competitive with other remote coupling schemes, such as ion shuttling or photonic interference~\cite{Stephenson2020-entanglement}, our wire-mediated coupling rate must also increase by two orders of magnitude to $\kappa \sim 2 \pi \times 1$~kHz. Following Eq.~\ref{eq:equiv_circuit_g}, we may achieve linear enhancement of the coupling rate by increasing the total number of ions~\cite{Harlander2011} or by lowering the resonant coupling frequency. Any further improvements will require additional engineering of the trap design and wire geometry, such as canceling some of the wire capacitance with an inductor~\cite{Heinzen1990}.

Ultimately, a well-developed wire-mediated coupling architecture may become a valuable tool for quantum computation and spectroscopy. At the single quanta level, this coupling allows deterministic entanglement of separately trapped ions via a simple solid-state link. Our work also shows potential for the development of trapped electron qubit architectures and hybrid quantum systems, where this coupling mechanism may serve as a quantum bridge between trapped ion qubits and superconducting qubits~\cite{Heinzen1990, Tian2004,Daniilidis2013electron}. 



\pagebreak
\bibliography{references}

\end{document}